\documentclass{svproc}
\usepackage{graphicx}
\usepackage{amsmath,amssymb,amsfonts}
\usepackage{wasysym}
\usepackage{afterpage}
\usepackage{mathpazo}
\usepackage{overpic}
\usepackage{multicol}
\usepackage{subfigure}
\usepackage{xspace}
\usepackage{rviewport}
\usepackage{colortbl}
\usepackage[hyphens]{url}
\usepackage[colorlinks=true,linkcolor=firebrick,citecolor=darkgreen,urlcolor=darkblue]{hyperref}
\usepackage{booktabs}
\usepackage[margin=1.5ex,bf]{caption}
\usepackage{a4wide}
\usepackage{color}
\usepackage{lineno} 
\usepackage{wrapfig}
\usepackage{cite}
\usepackage{multirow}
\usepackage{enumerate}
\usepackage[toc,page]{appendix}

\usepackage{blindtext}

\usepackage{tikz}

\usepackage{colortbl}
\usepackage{enumerate}
\usepackage{listings}
\usepackage{appendix}
\usepackage[final]{pdfpages}
\usepackage[Symbol]{upgreek}
\usepackage{makecell}

\graphicspath{{plots/}}
\definecolor{darkred}{rgb}{0.5,0,0}
\definecolor{darkblue}{rgb}{0,0,0.5}
\definecolor{firebrick}{rgb}{0.75,0.125,0.125}
\definecolor{darkgreen}{rgb}{0,0.5,0}

\usepackage[colorlinks=true,linkcolor=firebrick,citecolor=darkgreen,urlcolor=darkblue]{hyperref}

\usepackage{url}

\newcommand{\eV}{\ensuremath{\mbox{e\kern-0.1em V}}\xspace}
\newcommand{\GeV}{\ensuremath{\mbox{Ge\kern-0.1em V}}\xspace}
\newcommand{\MeV}{\ensuremath{\mbox{Me\kern-0.1em V}}\xspace}
\newcommand{\GeVc}{\ensuremath{\mbox{Ge\kern-0.1em V}\!/\!c}\xspace}
\newcommand{\GeVcc}{\ensuremath{\mbox{Ge\kern-0.1em V}\!/\!c^2}\xspace}
\newcommand{\AGeV}{\ensuremath{A\,\mbox{Ge\kern-0.1em V}}\xspace}
\newcommand{\AGeVc}{\ensuremath{A\,\mbox{Ge\kern-0.1em V}\!/\!c}\xspace}
\newcommand{\MeVc}{\ensuremath{\mbox{Me\kern-0.1em V}/c}\xspace}

\newcommand{\pt}{\ensuremath{p_{\textrm T}}\xspace}

\newcommand{\mt}{\ensuremath{m_{\textrm T}}\xspace}

\newcommand{\pim}{\ensuremath{\pi^-}\xspace}
\newcommand{\pip}{\ensuremath{\pi^+}\xspace}

\newcommand{\Xim}{\ensuremath{\Xi^-}\xspace}
\newcommand{\Xip}{\ensuremath{\overline{\Xi}^+}\xspace}

\newcommand{\Urqmd}{{\scshape U}r{\scshape qmd}\xspace}

\newcommand{\EposLong}{{\scshape Epos1.99}\xspace}

\newcommand{\CernVM}{\textsc{Cern\-\kern-0.05emVM}\xspace}

\newcommand{\TeV}{\ensuremath{\mbox{Te\kern-0.1em V}}\xspace}
\def \NASixtyOne{NA61/SHINE\xspace}

\begin{document}
\mainmatter              
\title{Recent results from NA61/SHINE}
\titlerunning{Recent results from NA61/SHINE}  
%
\author{Szymon Pulawski\inst{1} \\for the NA61/SHINE Collaboration}
\authorrunning{Szymon Pulawski} 
%
\tocauthor{Szymon Pulawski}
\institute{Institute of Physics, University of Silesia, Katowice, Poland
}

\maketitle              

\begin{abstract}
The research programme of the NA61 collaboration covers a wide range of
hadronic physics in the CERN SPS energy range, encompassesing measurements of hadron-hadron, 
hadron-nucleus as well as nucleus-nucleus collisions. The latter are analysed to better
understand the properties of hot and dense nuclear matter. In this paper recent results of particle production properties as well event by event fluctuations in proton-proton, Be+Be and Ar+Sc interactions at beam 
energies of 19\textit{A}/20\textit{A}, 30\textit{A}, 40\textit{A}, 75\textit{A}/80\textit{A} and 158\textit{A}~GeV/c are presented.
\end{abstract}

\section{The NA61/SHINE facility}

The NA61/SHINE detector~\cite{Abgrall:2014fa} is a large acceptance
hadron spectrometer with excellent capabilities in charged particle momentum measurements and
identification by a set of eight Time Projection Chambers as well as Time-of-Flight detectors. 
The high resolution forward calorimeter, the Projectile Spectator Detector (PSD), measures
energy flow around the beam direction, which in nucleus-nucleus reactions is primarily a measure
of the number of projectile spectator (non-interacted) nucleons and is thus related to the violence
(centrality) of the collision. A set of beam detectors identifies beam particles and measures precisely their trajectories.

NA61/SHINE performed a two-dimensional scan in collision energy (13\textit{A}-150\textit{A}~GeV/c and system size (p+p, Be+Be, Ar+Sc, Xe+La, Pb+Pb) to study the phase diagram of strongly interacting matter. The main goals of NA61/SHINE are the search for the critical point and a study of the onset of deconfinemnet.

\section{Study of the onset of deconfinement}

\subsection{Particle production properties}

The Statistical Model of the Early Stage (SMES)~\cite{Gazdzicki:1998vd} predicts a 1st order phase transition from the QGP to a hadron matter phase between top AGS and top SPS energies. In the transition region constant temperature and pressure in the mixed phase and an increase of the number of internal degrees of freedom is expected.

\begin{figure}[!ht]
\centering
\includegraphics[width=0.45\textwidth,clip]{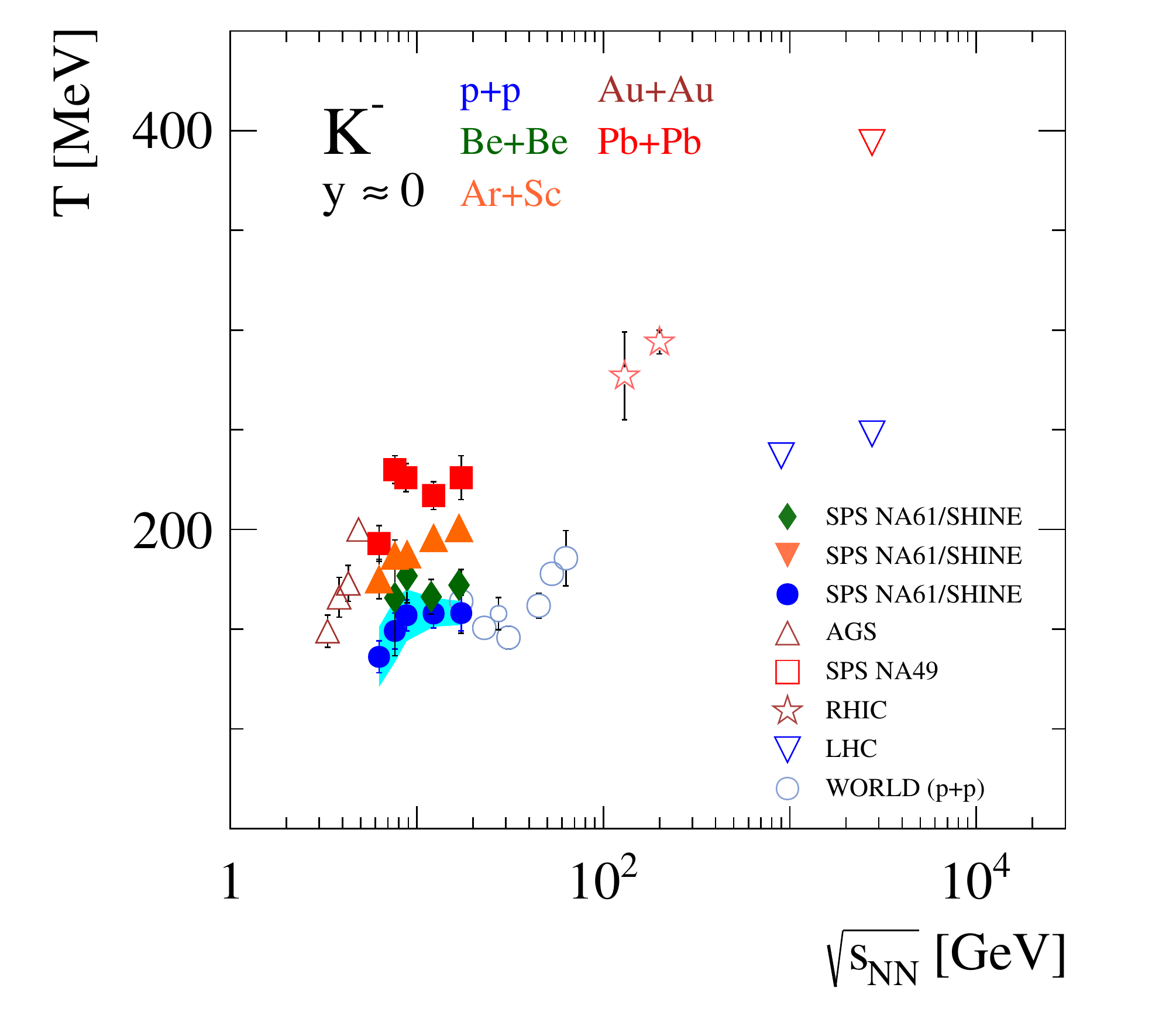}
\includegraphics[width=0.45\textwidth,clip]{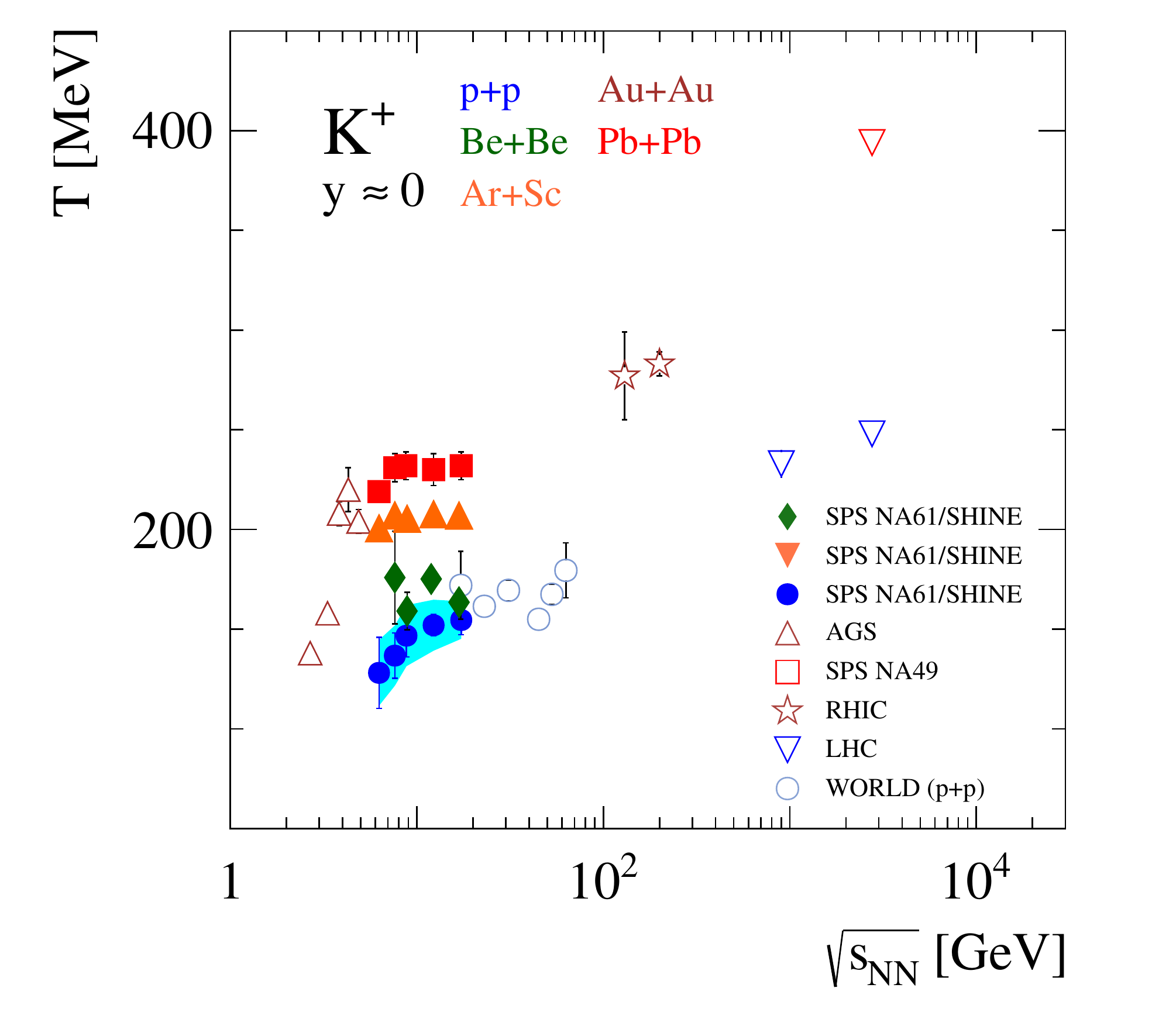}
\caption{Inverse slope parameter T of $m_{T}$ spectra of $K^{\pm}$ as function of collision energy. Most results are shown with statistical uncertainties only. For the p+p data the shaded band indicates systematic uncertainties.}
\label{fig:step}       
\end{figure}

A plateau ("step") in the energy dependence of the inverse slope parameter T was observed by the NA49 experiment in Pb+Pb collisions for $m_{T}$ spectra of $K^{\pm}$. It was expected for the onset of deconfinement due to the presence of a mixed phase of hadron gas (HRG) and quark-gluon plasma (QGP). In p+p interactions at SPS energies the inverse slope parameter T of $m_{T}$ spectra shows qualitatively similar enery dependence as in central Pb+Pb collisions ("step") and such a behaviour seems to emerge also in Be+Be reactions, as visible in Fig.~\ref{fig:step}. The values of the T parameter in Be+Be collisions are slightly above those in p+p interactions. The T parameter in Ar+Sc reactions is found between those in p+p/Be+Be and Pb+Pb collisions.

\begin{figure}[!ht]
\centering
\includegraphics[width=0.45\textwidth,clip]{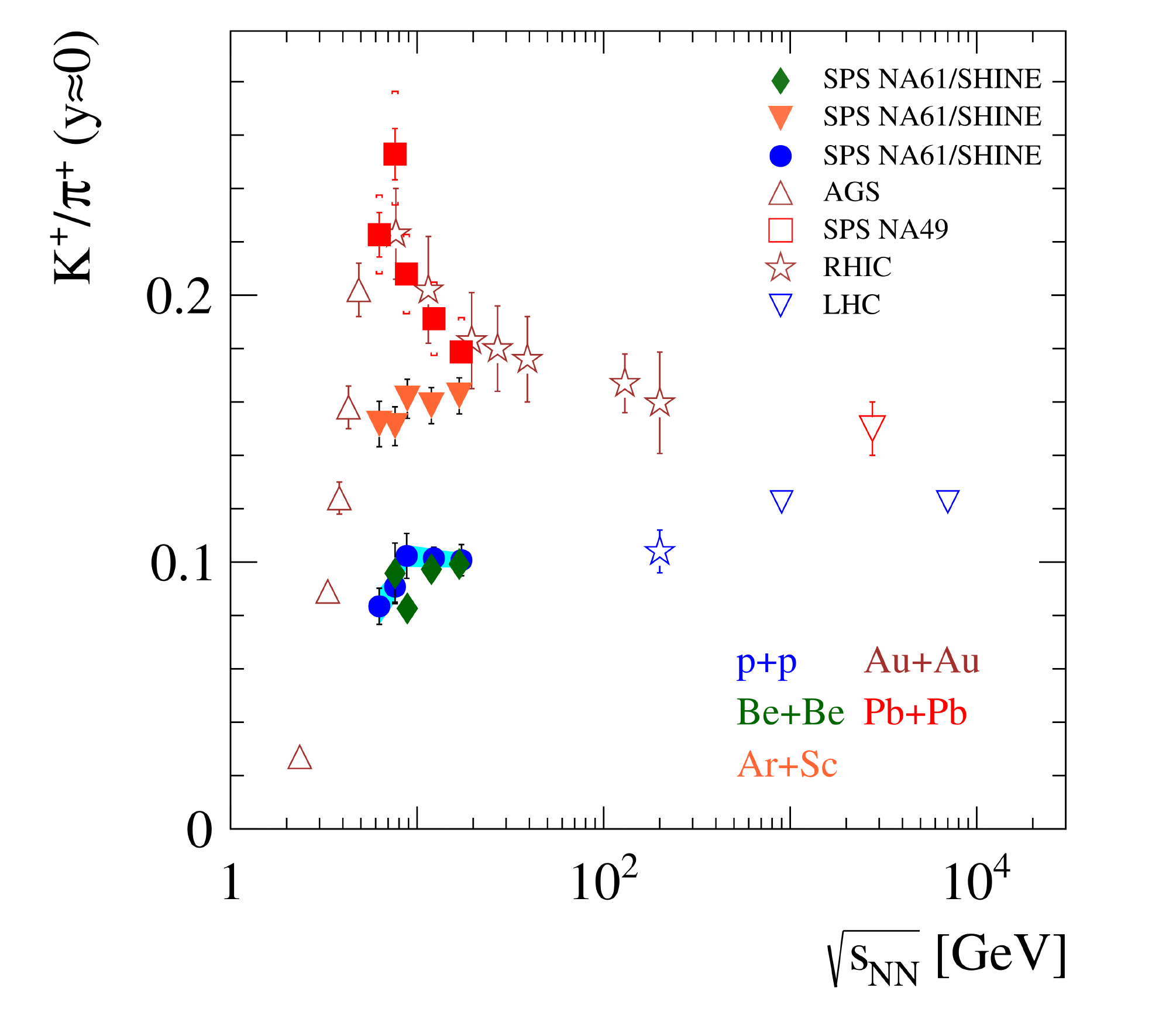}
\includegraphics[width=0.45\textwidth,clip]{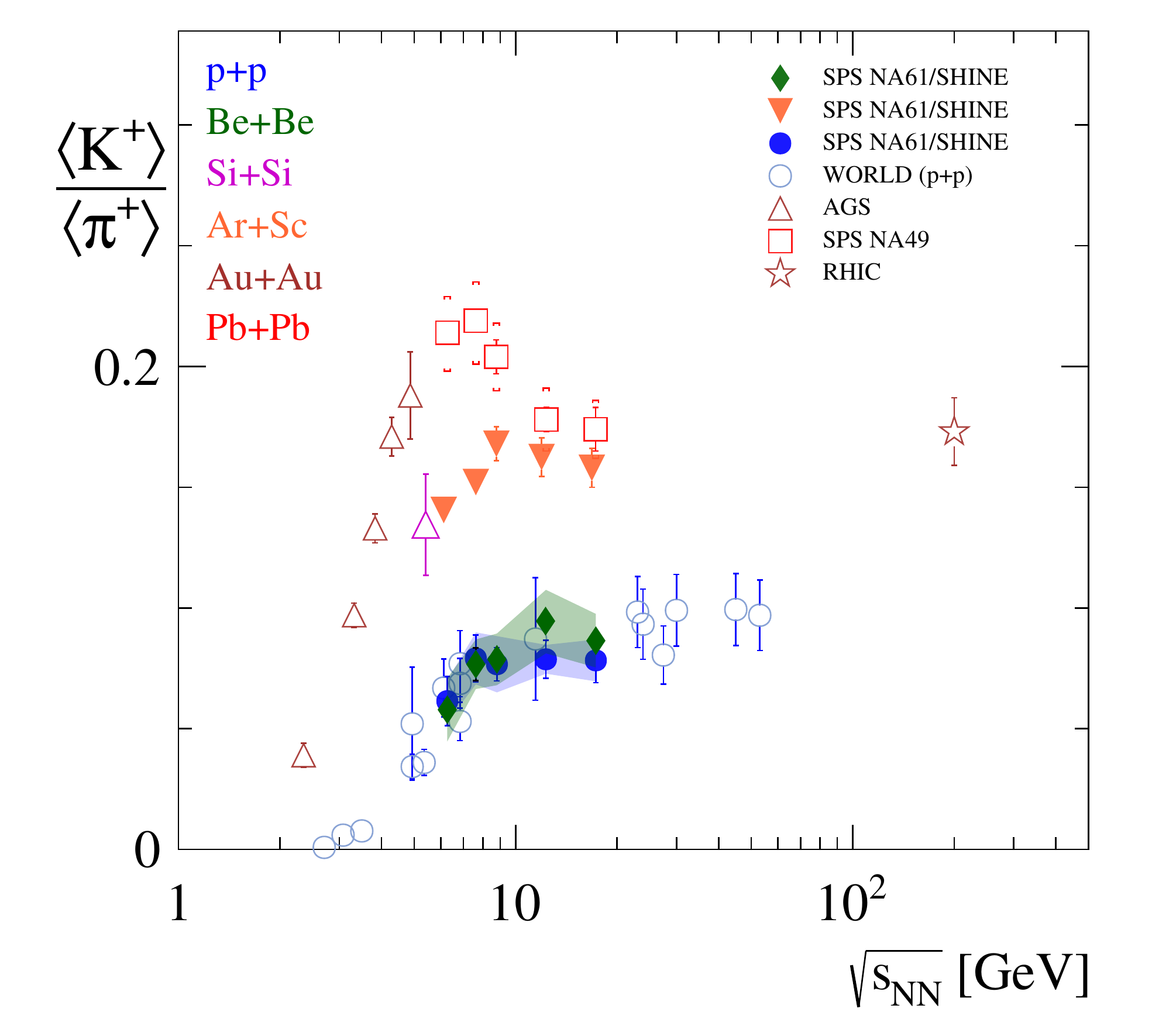}
\caption{Ratio of yields $K^{+}/\pi^{+}$ at mid-rapidity and the ratio of total yields $\left\langle K^{+} \right\rangle / \left\langle \pi^{+} \right\rangle$ produced in p+p, Be+Be and Pb+Pb collisions as function of collision energy.}
\label{fig:horn}       
\end{figure}

Finally, rapid changes of the ratios $K^{+}/\pi^{+}$ at mid-rapidity and $\left\langle K^{+} \right\rangle / \left\langle \pi^{+} \right\rangle$ as function of collision energy ("horn") were observed in Pb+Pb collisions by the NA49 experiment. These were predicted by the SMES model as a signature of the onset of deconfinement. These two ratios together with new NA61/SHINE results from Be+Be and Ar+Sc collisions are shown in Fig.~\ref{fig:horn}. A plateau like structure is visible in p+p interactions. The ratio $K^{+}/\pi^{+}$ at mid-rapidity as well as the ratio of total yields from Be+Be collisions is close to the p+p measurements. For the five analysed energies of Ar+Sc collisions, the ratio $K^{+}/\pi^{+}$ at mid-rapidity and $\left\langle K^{+} \right\rangle / \left\langle \pi^{+} \right\rangle$ are higher than in p+p collisions but show a qualitatively similar energy dependence - no horn structure visible.

\subsection{Flow}

Directed flow $v_1$ was considered to be sensitive to the first order phase transition (strong softening of the Equation of State)~\cite{Csernai:1999nf,Stoecker:2004qu,Brachmann:1999xt}. The expected effect is a non-monotonic behaviour (change from positive to negative and again to positive values) of proton $dv_1/dy$ as a function of beam energy. This effect is usually referred to as collapse of proton flow.  The NA49 experiment measured anti-flow of protons at mid-rapidity~\cite{Alt:2003ab}. A negative value of $dv_1/dy$ was observed in peripheral Pb+Pb collisions at 40$A$ GeV/c beam momentum ($\sqrt{s_{NN}}=8.8$ GeV). 

In 2018 the NA61/SHINE experiment reported the first results on anisotropic flow, measured in centrality selected Pb+Pb collisions at 30$A$ GeV/c beam momentum ($\sqrt{s_{NN}}=7.6$ GeV). According to the \textit{horn} structure in the energy dependence of the $K^{+}/\pi^{+}$ ratio in Pb+Pb collisions, this is the energy of the onset of deconfinement. Therefore, studying the centrality dependence of flow at this specific energy may allow to better understand the properties of the onset of deconfinement.

\begin{figure}
\centering
\includegraphics[width=0.5\textwidth]{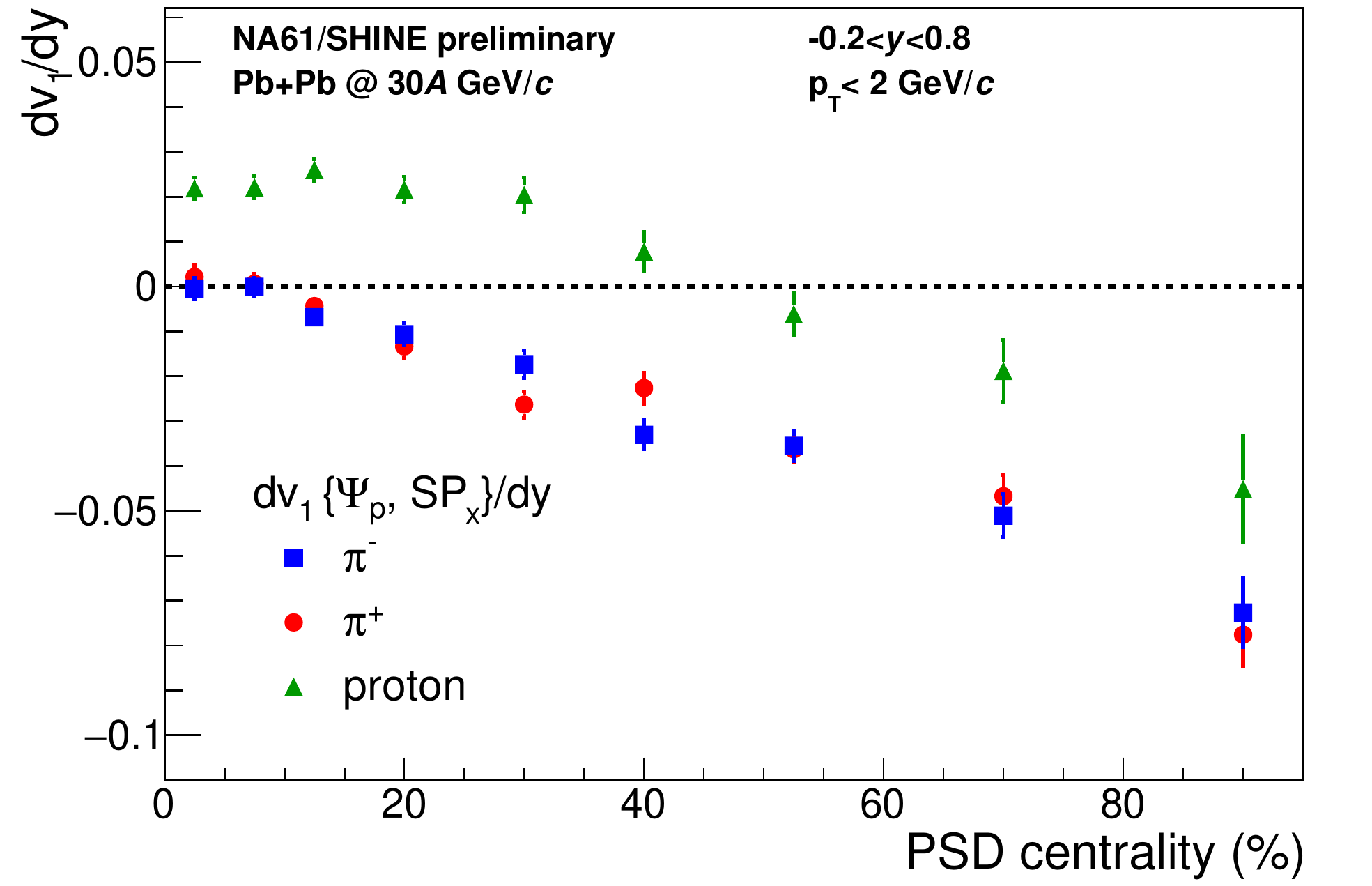}\includegraphics[width=0.33\textwidth]{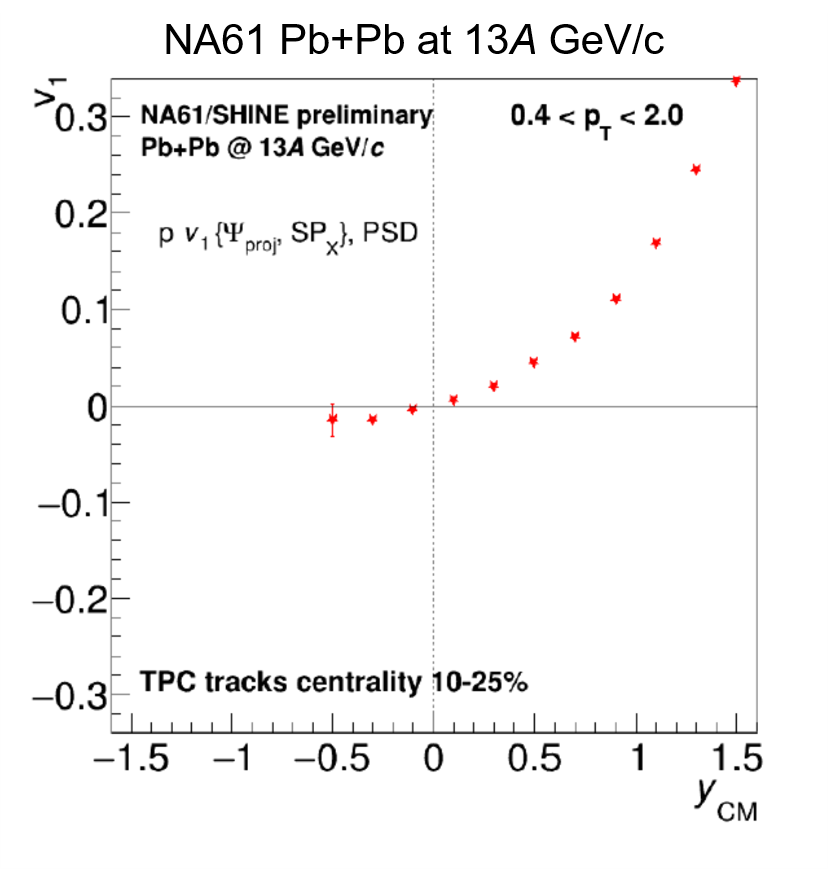}
\caption[]{\footnotesize Preliminary results on centrality dependence of $dv_1/dy$ at mid-rapidity measured in Pb+Pb collisions at 30$A$ GeV/c (\textit{left})and $v_1$ as the function of rapidity measured in Pb+Pb at 13$A$~GeV/c~(\textit{right}).}
\label{fig:dv1dy}
\end{figure}

The NA61/SHINE fixed target setup allows tracking and particle identification over a wide rapidity range. Flow coefficients were measured relative to the spectator plane estimated with the Projectile
Spectator Detector~(PSD), which is unique for NA61. Preliminary results on the centrality dependence of $dv_1/dy$ at mid-rapidity, measured in Pb+Pb collisions at 30$A$ GeV/c, are presented in Fig.~\ref{fig:dv1dy}~(\textit{left}). One sees that the slope of pion $v_1$ is always negative. In contrast, the slope of proton $v_1$ changes sign for centrality of about 50\%. Recently, preliminary results of directed flow for Pb+Pb collisions at 13$A$ were released~\cite{Klochkov:2018xvw}. Proton directed flow as function of rapidity is shown in Fig.~\ref{fig:dv1dy}~(\textit{right}). The results do not show evidence for the collapse of proton directed flow in Pb+Pb interactions at 13$A$~GeV/c.

\section{Search for the critical point}

An intermittency signal in protons was predicted close to the critical point. This is expected to become manifest in local power-law fluctuations of the baryon density which can be searched for by studying the scaling behaviour of second factorial moments $F_2(M)$ with the cell size or, equivalently, with the number of cells in ($p_x$, $p_y$) space of protons at mid-rapidity (see Refs.~\cite{Bialas:1985jb,Turko:1989dc,Diakonos:2006zz}). The transverse momentum space is partitioned into $M \times M$ equal-size bins, and the proton distribution is quantified by multiplicities in individual bins. The second order factorial moment in transverse momentum space is expressed as:

\begin{equation}
F_2(M)=\frac{\langle \frac{1}{M^2} \sum _{m=1}^{M^2}n_m(n_m-1) \rangle} 
{{\langle \frac{1}{M^2} \sum _{m=1}^{M^2}n_m \rangle}^2},
\end{equation}
where $M^2$ is the number of bins ($M$ bins in $p_x$ and $M$ bins in $p_y$) and $n_m$ is the number of protons in the $m$-th bin. Combinatorial background subtracted (by mixed events) second factorial moments, $\Delta F_2(M)$, should scale according to a power-law (for $M \gg 1$):
\begin{equation}
\Delta F_2(M) \sim (M^2)^{\phi_2}
\end{equation}  

\begin{figure}
\centering
\includegraphics[width=0.8\textwidth]{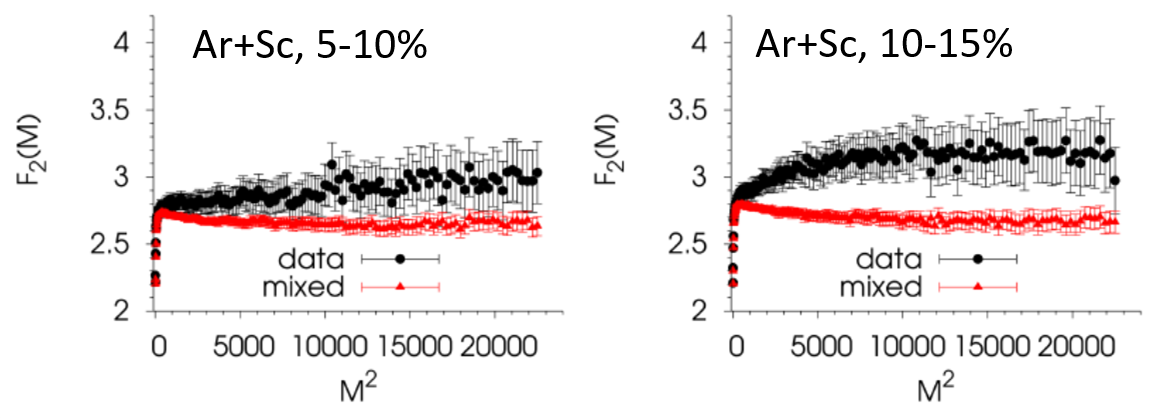}
\vspace{-0.2cm}
\caption[]{\footnotesize Preliminary results on $F_2(M)$ of mid-rapidity protons measured in 5-10\% (\textit{left}) and 10-15\% (\textit{right}) central Ar+Sc collisions at 150$A$ GeV/c.}
\label{fig:intermit_arsc}
\end{figure}

In the recent analysis of NA61/SHINE the intermittency effects were studied in central Be+Be and centrality selected Ar+Sc collisions at 150$A$ GeV/c. The $dE/dx$ method was used for the identification of protons. Centrality was determined from the energy deposited in the PSD detector. For Ar+Sc collisions protons were selected with at least 90\% purity. Figure~\ref{fig:intermit_arsc} shows preliminary results on $F_2(M)$ of mid-rapidity protons produced in 5-10\% and 10-15\% central Ar+Sc collisions at 150$A$ GeV/c.

The result of $F_2(M^2 )$seen in Ar+Sc collisions are higher in data than in mixed events. A detailed investigation of the significance of this result is in progress. 
 
A critical point is also expected to lead to enhanced fluctuations of multiplicity and transverse momentum. For their study NA61/SHINE uses the \textit{strongly intensive} measures $\Delta[P_T, N]$ and $\Sigma[P_T, N]$, see Ref.~\cite{Aduszkiewicz:2015jna}. In the Wounded Nucleon Model (WNM) they depend neither on the number of wounded nucleons ($W$) nor on fluctuations of $W$. In the Grand Canonical Ensemble they do not depend on volume and volume fluctuations. Moreover, $\Delta[P_T, N]$ and $\Sigma[P_T, N]$ have two reference values, namely they are equal to zero in case of no fluctuations and one in case of independent particle production. 

\begin{figure}
\centering
\includegraphics[width=0.8\textwidth]{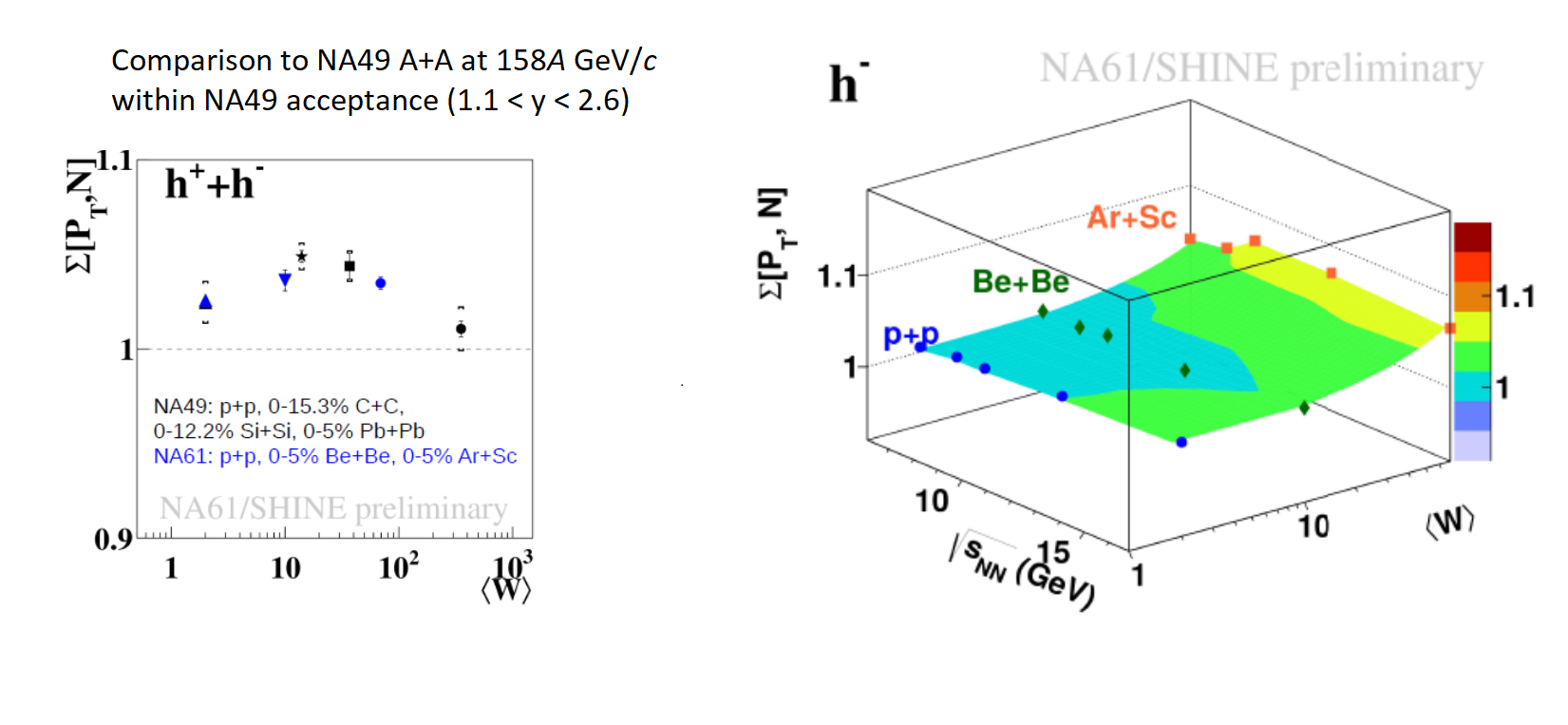}
\caption[]{$\Sigma[P_T, N]$ for all charged hadrons ($h^{+}+h^{-}$) from the NA61/SHINE and NA49 experiments as function of system size at 150$A$/158$A$ GeV/$c$ (\textit{left}) and $\Sigma[P_T, N]$ for negatively charged hadrons in inelastic p+p (blue squares), 0-5\% Be+Be (green diamonds), and 0-5\% Ar+Sc (orange squares) collisions obtained by NA61/SHINE. for NA61/SHINE only statistical uncertainties are shown. All NA61/SHINE results are preliminary.}
\label{fig:fluc}
\end{figure}

The system size dependence of $\Sigma[P_T, N]$ at 150$A$/158$A$ GeV/$c$ from the NA61/SHINE and NA49 experiments as function of system size (wounded nucleons) is shown in Fig.~\ref{fig:fluc} (\textit{left}). NA49 and NA61/SHINE measurements show consistent trends. Finally NA61/SHINE results for the NA61/SHINE acceptance for p+p, Be+Be and Ar+Sc collisions are presented in Fig.~\ref{fig:fluc} (\textit{right}). So far there are no prominent structures observed which could be related to a critical point.

\section{Strangeness production in p+p interactions at 158 GeV/c}

\subsection{$\Xi$ production}

Hyperons are excellent probes of the dynamics of proton-proton interactions as constituent strange quarks are
not present in the initial state of this process. Therefore hyperon production has been studied in a long series of experiments in elementary hadron+hadron interactions. However, the experimental situation in this field remains inconclusive.

New data from p+p collisions on ${\Xi}^{-}$ and $\overline{\Xi}^{+}$ hyperon production are presented. The event sample consists of 53 million registered interaction trigger events obtained at 158~\GeVc beam momentum corresponding to $\sqrt{s_{NN}}$ = 17.3~\GeVc. The results refer to primary ${\Xi}^{-}$ and $\overline{\Xi}^{+}$ produced in strong and electromagnetic processes and are corrected for detector geometrical acceptance and reconstruction efficiency.

\begin{figure}[ht]
	\centering
 	\includegraphics[width=0.45\textwidth]{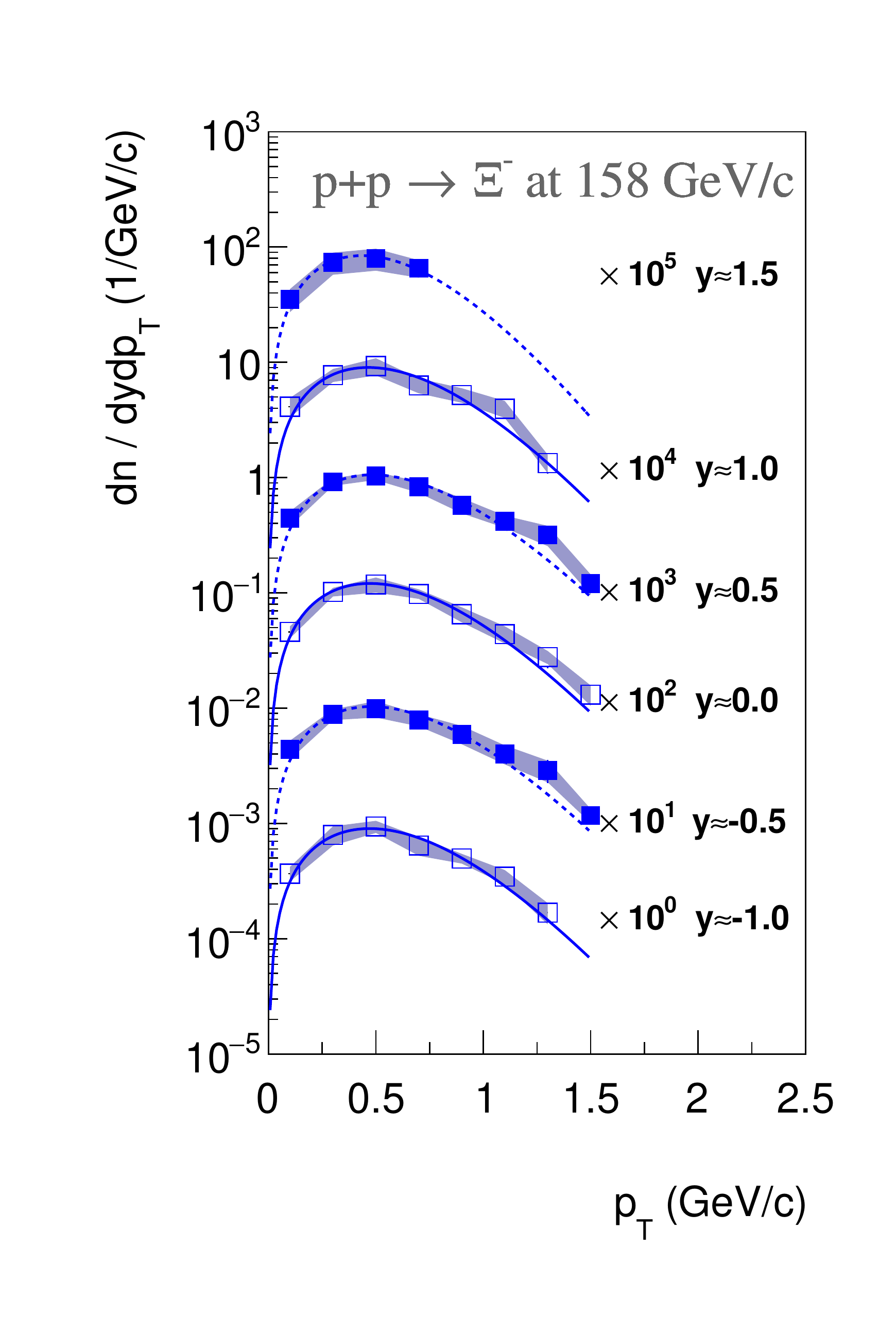}
 	\includegraphics[width=0.45\textwidth]{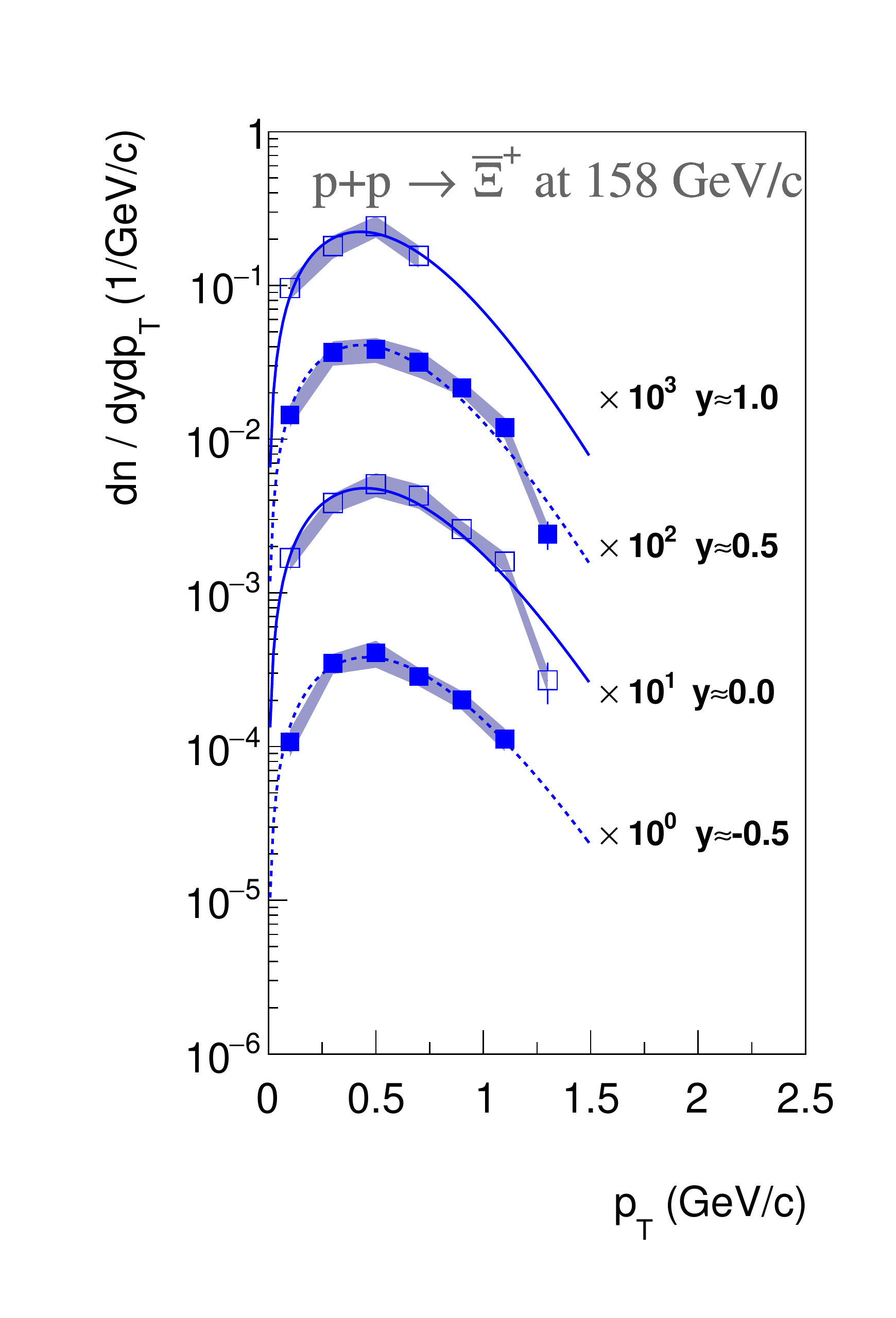}
	\vspace{-0.7cm}
	\caption{Preliminary results on transverse momentum spectra of $\Xi^{-}$ (\textit{left}) and  $\overline{\Xi}^{+}$~(\textit{right}) hyperons produced in inelastic p+p interactions at 158~\GeVc in consecutive rapidity bins. Results are scaled for better separation, shaded bands indicate systematic uncertainty.}
	\label{fig:Xi_rapidity}
\end{figure}

To find the $\Xi$ candidates, all $\Lambda$ candidates are combined with pion tracks of appropriate charge (daughter track). A fitting procedure is applied, using as parameters the decay position of the $V^{0}$
candidate, the momenta of both the $V^{0}$ decay tracks, the momentum of the daughter track, and finally the $z$
position of the $\Xi$ decay point. The $x$ and $y$ position of the $\Xi$ decay position are not subject to the minimization, as they are determined from the parameters using momentum conservation. This procedure yields the decay position and the momentum of the $\Xi$ candidate.

\begin{figure}[ht]
	\centering
	\includegraphics[width=0.45\textwidth]{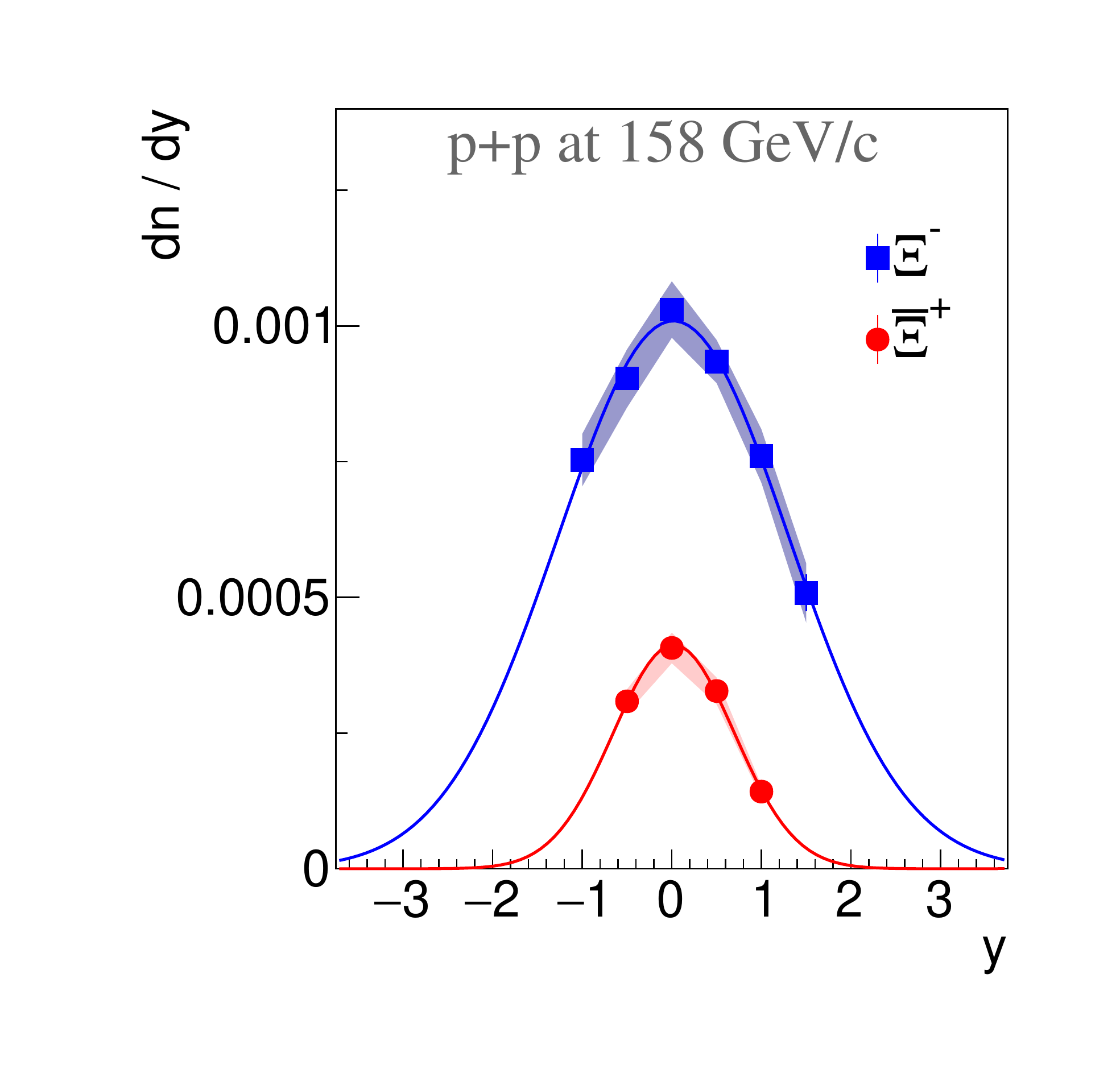}
	\vspace{-1cm}
	\caption{Preliminary results on rapidity spectrum of $\overline{\Xi}^{+}$ (red) and $\Xi^{-}$ (blue) hyperons produced in inelastic p+p interactions at 158~\GeVc fitted by a Gaussian function (lines).}
	\label{fig:Xidndy}
\end{figure}

Preliminary results derived from two dimensional spectra ($y$ vs \pt) are presented as transverse momentum distributions in bins of rapidity in Fig.~\ref{fig:Xi_rapidity}. Statistical uncertainties are shown as vertical bars and preliminary estimates of systematic uncertainty are indicated by shaded bands. The blue lines show results of exponential fits to the measurements binned in \mt.

The obtained \pt spectra were used to calculate the rapidity spectrum of $\Xi^{-}$ and $\overline{\Xi}^{+}$ production as the sum of measured points and extrapolation to the unmeasured region of \pt. The result is displayed in Fig.~\ref{fig:Xidndy}. Vertical bars show statistical, the shaded band systematic uncertainties. The rapidity distribution was fitted by a Gaussian function for extrapolation into the unmeasured regions. Based on summing the data points and the extrapolation of the fitted function resulted in the mean multiplicity $\langle \overline{\Xi}^{+} \rangle = $ 0.00079 $\pm$ 0.00002 $\pm$ 0.00010 and the mean multiplicity of $\langle \Xi^{-} \rangle =$  0.0033 $\pm$ 0.0001 $\pm$ 0.0006.

\begin{figure}[ht]
	\centering
 	\includegraphics[width=0.45\textwidth]{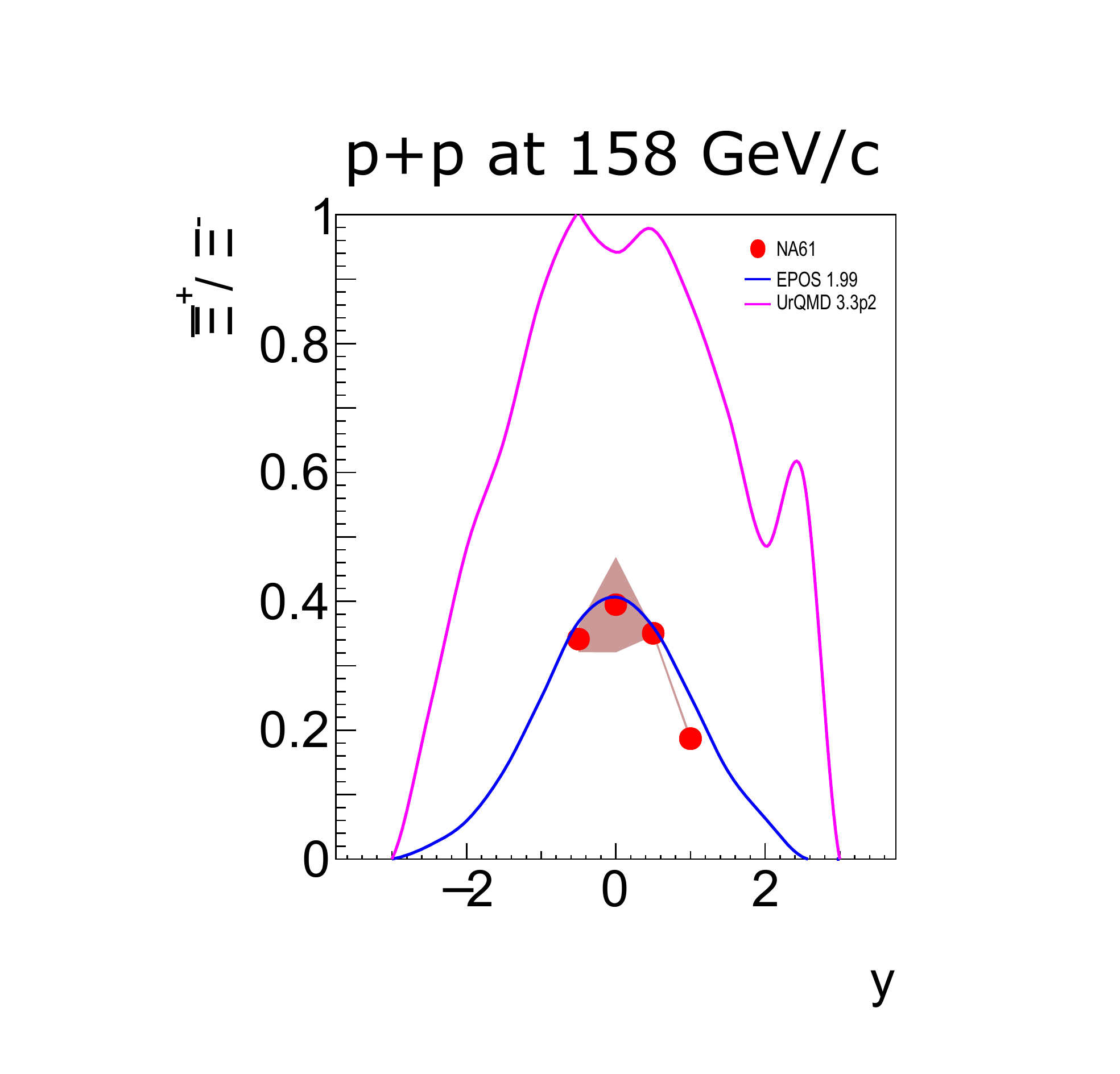}
	\vspace{-0.6cm}
	\caption[]{Ratio of rapidity spectra $\overline{\Xi}^{+}/{\Xi}^{-}$  compared with \Urqmd (magenta solid line) and \EposLong (blue solid line) models predictions.}
	\label{fig:Ximodely}
\end{figure}

Additionally, the ratio of rapidity spectra $\overline{\Xi}^{+}/{\Xi}^{-}$  was calculated and compared with \Urqmd and \EposLong model predictions in Fig.~\ref{fig:Ximodely}. \Urqmd fails to describe $\overline{\Xi}^{+}/{\Xi}^{-}$, which is a known problem of string models. \EposLong describes the rapidity distributions of $\overline{\Xi}^{+}$, ${\Xi}^{-}$ and their ratio, but not the shape of the transverse momentum spectra.

\subsection{ Search for pentaquark candidates}

The NA49 Collaboration published evidence for the existence of a narrow \Xim\pim baryon resonance with mass of 1.862 $\pm$ 0.002~\GeVcc and width below the detector resolution~\cite{Alt:2003vb} in 2004. The significance was estimated to be 4.0$\sigma$. This state was a candidate for the hypothetical exotic  $\Xi^{--}_\frac{3}{2}$ baryon with S=-2, I=$\frac{3}{2}$ and a quark content of ($dsds\bar{u}$). At the same mass a peak was observed in the \Xim\pip spectrum which is a candidate for the $\Xi^{0}_\frac{3}{2}$ member of this isospin quartet with a quark content of ($dsus\bar{d}$). The corresponding antibaryon spectra also showed enhancements at the same invariant mass.

Recently a similar analysis was performed by \NASixtyOne based on an order of magnitude higher statistics. The first step in the analysis was the search for $\Lambda$ candidates, which were then combined with the \pim to form the \Xim candidates. Next the $\Xi^{--}_\frac{3}{2}$ $\left(\Xi^{0}_\frac{3}{2}\right)$ were searched for in the \Xim\pim (\Xim\pip) invariant mass spectrum, where the \pim (\pip) are primary vertex tracks. An analogous procedure was followed for the antiparticles.

\begin{figure}[t!]
\begin{tikzpicture}
	\begin{scope} [xshift=-3.5cm]
	\node {\includegraphics[width=0.45\textwidth]{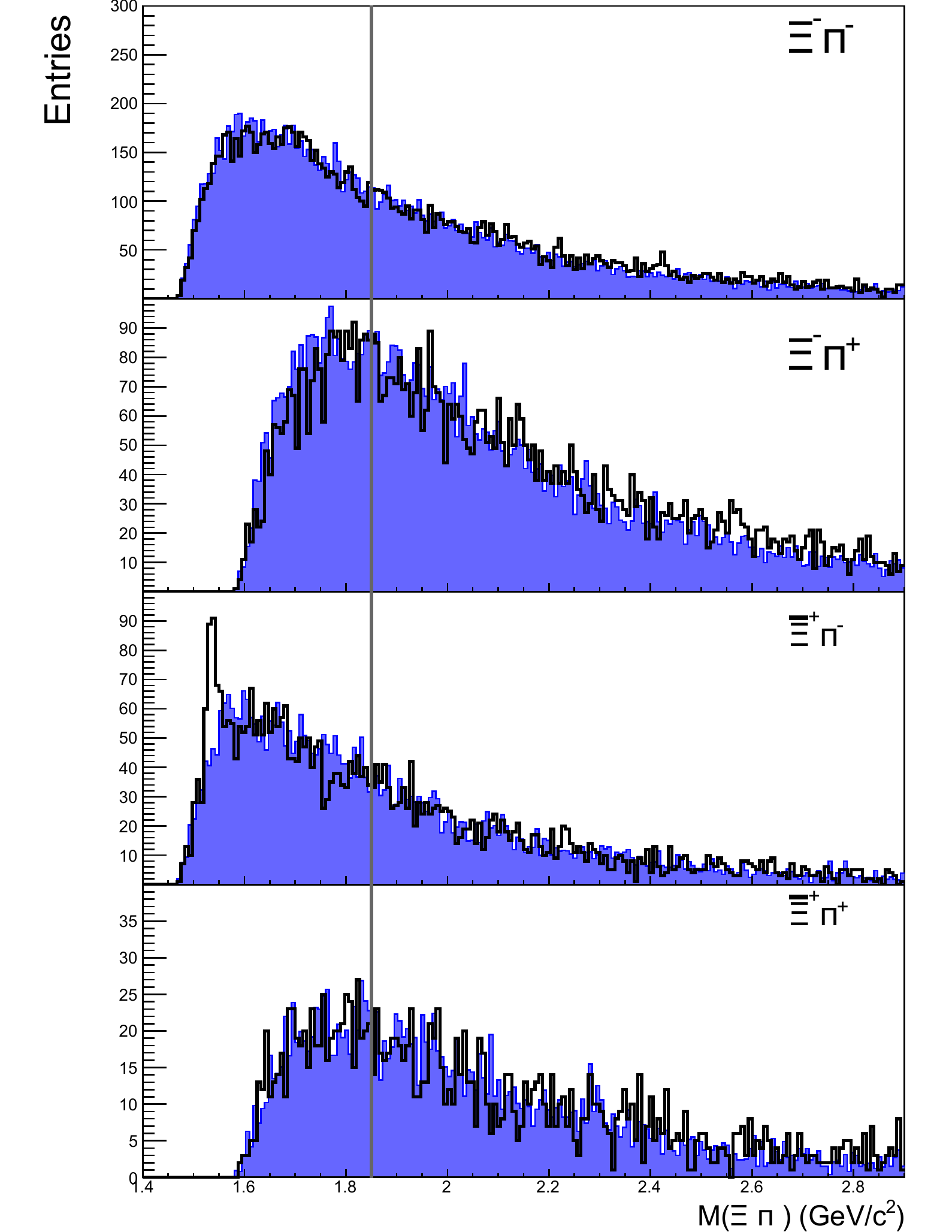} };
	\end{scope}
	\begin{scope} [xshift=4.0cm, yshift=-0.05cm]		
	\node {\includegraphics[width=0.525\textwidth]{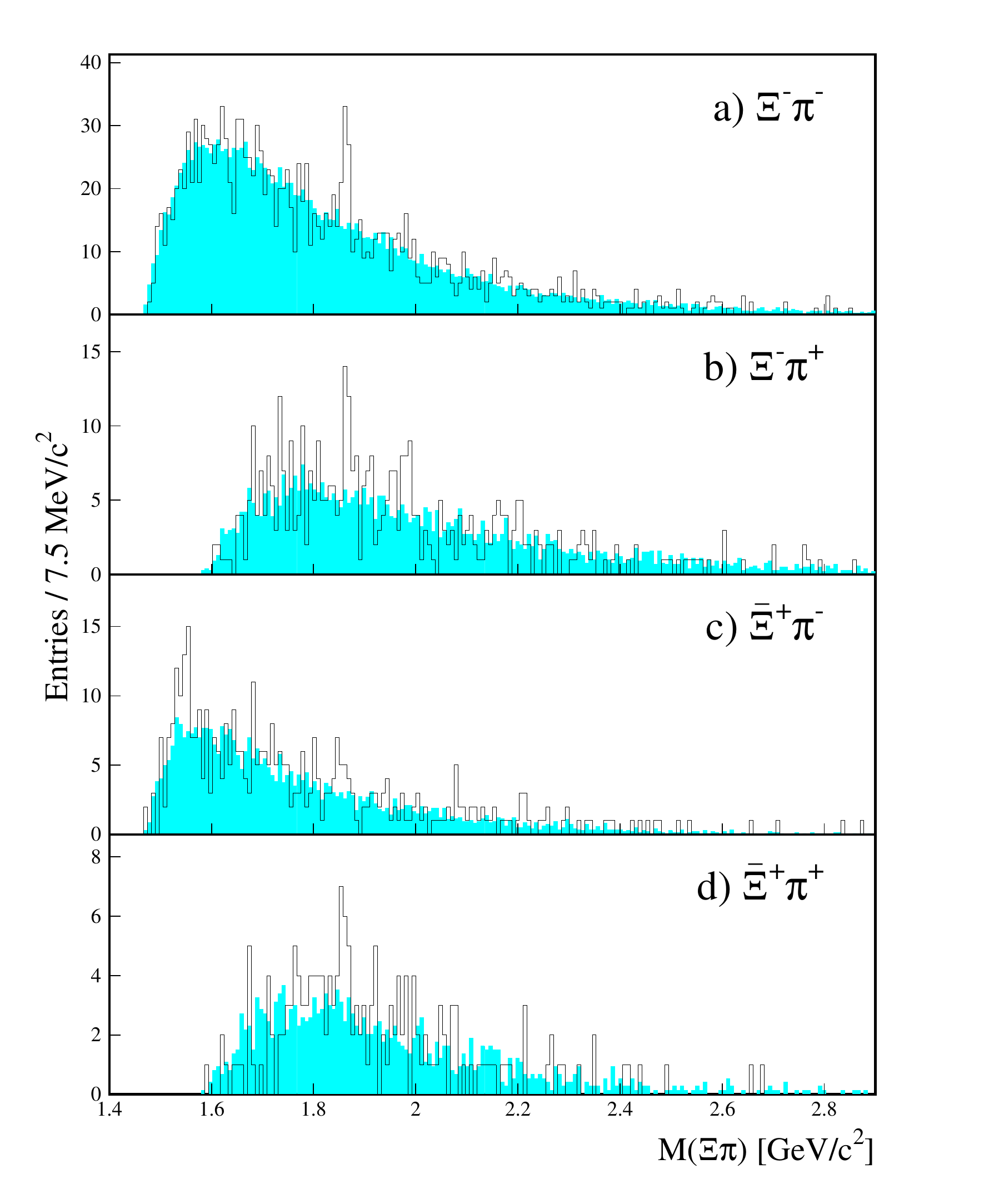} };
	\end{scope}
\end{tikzpicture}
\caption{\textit{Left}: \NASixtyOne invariant mass spectra after selection
cuts for \Xim\pim, \Xim\pip, \Xip\pim (note that the $\Xi(1530)^0$ state is also visible) and \Xip\pip. The shaded histograms are the normalised mixed-event backgrounds. Gray line point to place where indication of the signal was observed by the NA49 Collaboration. \NASixtyOne results were obtained for 53M interaction trigger events (33M after all event cuts). \textit{Right}: Similar plots but obtained by NA49~\cite{Alt:2003vb}. In NA49 6.5M recorded events (3.75M after event cuts) were used.  
}
\label{fig:penta}
\end{figure}

To search for the exotic $\Xi^{--}_\frac{3}{2}$ state the selected \Xim candidates were combined with primary \pim tracks. %
The resulting \Xim\pim invariant mass spectrum is shown in Fig.~\ref{fig:penta} (\textit{top, left}). The shaded histogram is the mixed-event background, obtained by combining the \Xim and \pim from different events and normalising to the number of real combinations. 
The complete set of invariant mass distributions measured by \NASixtyOne (\Xim\pim, \Xim\pip, \Xip\pim, \Xip\pip) is shown in Fig.~\ref{fig:penta} (\textit{left}). In addition to the described cuts, a lower cut of 3~\GeVc was imposed on the \pip momenta to minimize the large proton contamination. Blue histograms show normalised mixed-event backgrounds. One sees that data overlap with the mixed-event backgrounds in the mass window 1.848 -- 1.870~\GeVcc where the NA49 signal was observed; see Fig.~\ref{fig:penta} (\textit{right}). Finally a narrow peak of $\Xi(1530)^0$ is observed in the invariant mass of \Xip\pim. The yield of observed $\Xi(1530)^0$ scales with the number of events compared to NA49 results. 

In summary, this \NASixtyOne analysis of p+p interactions, with $\approx$ 10 times higher statistics, disproves the NA49 indication of the production of $\Xi^{--}_\frac{3}{2}$, $\Xi^{0}_\frac{3}{2}$ and their antiparticles. All four invariant mass distributions shown in Fig.~\ref{fig:penta} do not show significant signals in the mass window for which NA49 previously reported pentaquark candidates. 

\section*{Acknowledgments}
This work was supported by
the Polish Ministry of Science
and Higher Education (grants 667\slash N-CERN\slash2010\slash0,
NN\,202\,48\,4339 and NN\,202\,23\,1837), the National Science Centre Poland 2015\slash18\slash M\slash ST2\slash00125.

\bibliographystyle{include/na61Utphys}
{\footnotesize\raggedright
\bibliography{include/na61References}
}

\end{document}